\documentclass[aps,preprint,prd,epsfig,axodraw,nofootinbib]{revtex4-1}
\usepackage{amssymb}
\usepackage{amsmath}
\usepackage{graphicx}
\usepackage{fancyhdr}
\usepackage{epsfig}
\usepackage{axodraw}
\usepackage{color}
\usepackage[bookmarks=true]{hyperref}

\def\br{\begin{eqnarray}}
\def\er{\end{eqnarray}}
\def\be{\begin{equation}}
\def\ee{\end{equation}}

\def\g{\gamma}

\def\to{\rightarrow}

\def\({\left(}
\def\){\right)}

\def\gev{\; \hbox{GeV}}

\def\lesssim{\mathrel{\hbox{\rlap{\hbox{\lower4pt\hbox{$\sim$}}}\hbox{$<$}}}}
\def\gtrsim{\mathrel{\hbox{\rlap{\hbox{\lower4pt\hbox{$\sim$}}}\hbox{$>$}}}}
\begin{document}

\title{Explaining the Higgs Decays at the LHC with an Extended Electroweak Model}
%\title{Explaining the Higgs Decays at the LHC and Tevatron with an  Extended Electroweak Gauge Sector Model}
%\title{Extended Electroweak Gauge Sectors -- A Natural Explanation for the LHC and Tevatron Data on Higgs Physics}

\author{\vspace{0.5 cm} Alexandre Alves$^{1}$, E. Ramirez Barreto$^{2}$, A. G. Dias$^{2}$, \\C. A. de S. Pires$^3$, Farinaldo S. Queiroz$^{3,4}$, and P. S. Rodrigues da Silva$^{3}$\vspace{0.5 cm}}

%\author{A. Alves}
\affiliation{{$^1$Departamento de Ci\^encias Exatas e da Terra,
Universidade Federal de S\~ao Paulo,  Diadema-SP, 09972-270, Brasil.\vspace{0.5 cm}}}

%\author{E. Ramirez Barreto, A. G. Dias}
\affiliation{{$^2$Centro de Ci\^encias Naturais e Humanas, Universidade Federal do ABC,  Santo Andr\'e-SP, 09210-170, Brasil.\vspace{0.5 cm}}}

%\author{C. A. de S. Pires$^a$, F. S. Queiroz$^{a,b}$, and P. S. Rodrigues da Silva$^{a}$}

\affiliation{{$^3$Departamento de
F\'{\i}sica, Universidade Federal da Para\'\i ba, Caixa Postal 5008, 58051-970,
Jo\~ao Pessoa, PB, Brasil\vspace{0.2 cm}\\
$^4$Center for Particle Astrophysics, Fermi National Accelerator Laboratory, Batavia, IL 60510, USA
}}

\date{\today}

\begin{abstract}
We show that the observed enhancement in the diphoton decays of the recently discovered new boson at the LHC, which we assume to be a Higgs boson, can be  naturally explained by a new doublet of charged vector bosons from extended electroweak models with  SU(3)$_C\otimes$SU(3)$_L\otimes$U(1)$_X$ symmetry. These models are also rather economical in explaining the measured signal strengths, within the current experimental errors, demanding fewer assumptions and less parameters tuning. Our results show a good agreement between the theoretical expected sensitivity to a 126--125 GeV Higgs boson, and the experimental significance observed in the diphoton channel at the 8 TeV LHC. Effects of an invisible decay channel for the Higgs boson are also taken into account, in order to anticipate a possible confirmation of deficits in the branching ratios into $ZZ^*$, $WW^*$, bottom quarks, and tau leptons.
%\\
%PACS: 14.60.St; 14.60.Pq; 12.60.Cn; 12.60.Fr.
%
\end{abstract}

\maketitle

\section{ Introduction}
\label{intro}
It was announced recently at the CERN Large Hadron Collider (LHC) the discovery of a new boson  whose observed properties until now suggest it is the Standard Model (SM) Higgs boson~\cite{CMS-ATLAS-sem2012}. Denoting by $h$ such new boson, its observation was based on decay signals in four leptons, $h \rightarrow l^+l^-l^+l^-$,  and diphotons, $h\rightarrow\gamma\gamma$, with both pointing to an invariant mass of  $m_h=$125 -- 126 GeV. The diphoton channel points to an excess over what is expected from the SM. We shall consider in this work that the $h$ is indeed a Higgs boson, i. e., as resulting from spontaneous symmetry breaking~\cite{hebghk}.

Within the observed mass range, several decay channels for the SM Higgs boson are experimentally accessible making  possible the measurement of its coupling strength to many particles. It is not clear yet from the present data if $h$ has its couplings to fermions as dictated by the SM, even adding the latest results from Tevatron \cite{CDF-D0-sem2012} which indicate an excess of $b-jets$ events, probably due to the decay $h\rightarrow b\bar{b}$. We expect that this will soon be resolved with more accumulated data. If the tendency of smaller branching ratios of $h$ into $b$ quarks and $\tau$ leptons is confirmed, this could be a smoking gun for models with a fermiophobic Higgs or models with a decreased effective coupling of the Higgs with the gluons. On the other hand, if these couplings have the strength as in the SM but their branching ratios turn out to be smaller than the expected, then invisible decay channels may have an important role.

In a previous work~\cite{h2f331min} we investigated in what extension an excess for the diphoton channel can be used to probe new vector bosons within a specific framework of a class of SU(3)$_C\otimes$SU(3)$_L\otimes$U(1)$_X$ gauge models, minimal 331 model for short~\cite{ppf-tp}. Our updated results show a good agreement between our theoretical expected sensitivity to a 126--125 GeV Higgs boson and the experimental significance observed in the diphoton channel at the 8 TeV LHC.

Facing the new experimental reality we now present a new focus in the diphoton channel, showing how the observed excess of photons can be explained by a new SU(2)$_L$ doublet of vector bosons. Such a doublet is contained in the minimal 331 model but may be part of other models with an extended electroweak gauge sector as well.

We also have included an analysis of an invisible decay width for the Higgs boson, in order to anticipate a possible confirmation of deficits in the branching ratios into $Z$, $W$, bottom quarks, and tau leptons. This class of models is truly the most economical one in the sense that all the tree level couplings of the Higgs boson to the $Z$, $W$, and fermions are exactly the same as the SM ones at the same time its possible to enhance the effective 1-loop coupling to photons. Moreover, there are not new contributions to the effective coupling between the Higgs and the gluons, thus all the Higgs production rates coincides with the SM predictions.

The quest for a mechanism of enhancement in the diphoton channel, in accordance with the recent results of \cite{CMS-ATLAS-sem2012}, was treated in \cite{carena-etal, gunion-etal, arbeya-etal, benbrik-etal, chang, barger, bellazzini,cheung,cheon,montull} taking into account specific models. Several independent analyses indicate deviations from the SM expectations \cite{ellis-you, corbett-etal}, and also include an invisible branching decay rate~\cite{espinosa-etal, giardino-etal, carmi-etal, farinaldo-etal} in order to explain the discrepancies.

We found that our results are consistent with these works. Our analysis shows that a new $SU(2)_L$ doublet of charged vector bosons of masses $\approx 213\gev$  and an ${\cal O}$(10\%) branching ratio of the Higgs boson into invisible states can reasonably fit part of the available data released by the LHC and Tevatron collaborations on Higgs branching ratios.

\section{Higgs--Vector Bosons Interactions}
\label{Higgs--Vector Bosons Interactions}

The doublet of vector bosons we take into account here has hypercharge $Y=3$
\begin{equation}
{\cal{V}} = \begin{pmatrix}
U^{++}   \\
V^+
\end{pmatrix}\sim({\bf{2}}\,,\,3).
\label{vecdub}
\end{equation}
Let $m_W$ be  the $W$ boson mass and $m_V$ the mass of these new vector bosons, which we consider to be degenerated. As the new vector bosons masses are related with an energy scale $v_\chi$, above the electroweak scale $v_W$ = 246 GeV, their contribution to the  process amplitude is multiplied by a suppression factor $m_W^2/m_V^2$ in comparison with that one coming from the $W$ boson. But  there is still a significant increasing of the branching $Br(h\rightarrow\gamma\gamma)$ so that a signal above the SM is indeed observed for an interesting range of $m_V$. This is due a dominant contribution of vector gauge bosons, and the fact that a double charged one leads to a factor four multiplying the suppression factor.

Interactions of the Higgs boson field with the new vector bosons are described by the following interaction Lagrangian
\br
{\cal L}_{HVV} &=&  2(\sqrt{2}G_F)^\frac{1}{2}m_W^2\,h(c_U U^{++\mu} U_{\mu}^{--}+c_V V^{+\mu} V_{\mu}^-)\nonumber\\
& & + g^2\,h^2(U^{++\mu} U_{\mu}^{--}+V^{+\mu} V_{\mu}^-).
\label{lhv}
\er
where $G_F$ is the Fermi constant, with $c_U$ and $c_V$ coefficients of order one. We take here the specific configuration of the models in Ref. \cite{h2f331min} where $c_U=c_V=\frac{1}{2}$. Also, we checked that contributions due the couplings of the Higgs boson with additional charged scalars are small enough for being disregarded. In fact, a charged scalar with mass comparable with the vector bosons $U$ and $V$ can only give sub-dominant contributions.

\begin{figure}
\centering
\includegraphics[width=0.5\columnwidth]{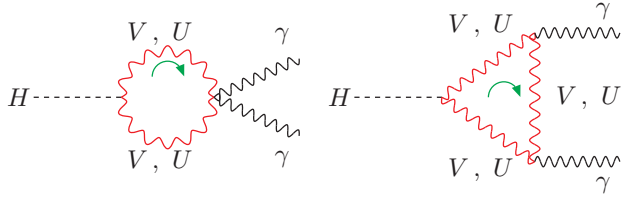}
\caption{The one-loop diagrams involving the new charged gauge bosons $U$ and $V$ which contribute most to the $h\to\gamma\gamma$ decay amplitude.}
\label{fig1}
\end{figure}

The diphoton Higgs boson decay is described by the effective Lagrangian
\br
{\cal L}_{H\g\g} &=& \frac{\alpha (\sqrt{2}G_F)^\frac{1}{2}}{8\pi} (F^{SM}+F^{new}) h F^{\mu\nu}F_{\mu\nu}\,.
\label{lefh2f}
\er
$\alpha$ is the fine structure constant, $F^{SM}$ and $F^{new}$ are structure function coefficients. $F^{SM}$ is what is obtained taking into account the interactions of the SM Higgs boson. The expression for $F^{SM}$ can be found in Refs.~\cite{Ellis,Okun,Hunters}.  All nonstandard couplings of the Higgs boson field with electrically charged fields gives rise to  $F^{new}$. In the case we are considering, this last coefficient is obtained from the trilinear interaction  in Eq. (\ref{lhv}) and new vector boson coupling with the photon. The Feynman diagrams involved  are shown in Figure ~\ref{fig1}.
The result of these diagrams can be obtained from the corresponding ones for $W$ boson \cite{Ellis,Okun,Hunters} just multiplying by a scale factor  proportional to $c_U m_W^2/m_V^2$. Thus, we have
\br
F^{new}= 5\left[2+3\tau_{V} +3\tau_{V}\left(2-\tau_{V}\right)I^2\right]\frac{m_{W}^2}{2m_{V}^2}\,,
\label{Fnew}
\er
with
\be
\tau_V \equiv \frac{4 m_V^2}{m_h^2}\,,
\ee
and
\be
I \equiv \left\{
\begin{array}{l}
arcsin\left(\sqrt{\frac{1}{\tau_{V}}}\right)\,\,\,\hbox{for} \,\,\,\tau_{V}\geq 1 \\
\frac{1}{2}\left[\pi+\imath \ln\left[\frac{1+\sqrt{1-\tau_{V}}}{1-\sqrt{1-\tau_{V}}}\right]\right]\,\,\,\hbox{for} \,\,\,\tau_{V}\leq 1
\end{array}\right.
\ee
For the diphoton the Higgs boson decay rate we then have,
\be
\Gamma_{h\g\g} = \frac{\alpha^2m_h^3 G_F}{128\sqrt{2}\pi^3}\left|F^{SM}+F^{new} \right|^2\,.
\label{largura}
\ee

\section{Observing a Higgs boson from 331 models with and without Dark Matter contents}

In 331 models, the light Higgs boson couples to the $SU(2)_L$ doublets of new gauge bosons and scalars but not to the new fermions, so we assume that the cross sections for the main light Higgs production processes are the same as the SM at the LHC and the Tevatron, i.e.
\begin{eqnarray}
\sigma_{331}(gg\to h) &=& \sigma_{sm}(gg\to h)\nonumber\\
\sigma^{vbf}_{331}\left(pp(\bar{p})\to hjj\right) &=& \sigma^{vbf}_{sm}\left(pp(\bar{p})\to hjj\right)\nonumber\\
\sigma_{331}\left(pp(\bar{p})\to hZ(W)\right) &=& \sigma_{sm}\left(pp(\bar{p})\to hZ(W)\right)
\label{prod}
\end{eqnarray}

The new charged gauge bosons mediate interactions between exotic and SM quarks only, that is why the 1-loop effective gluon-gluon-Higgs coupling does not receive new contributions. All the tree level couplings between the Higgs and  all the fermions, the $Z$, and  the $W$ bosons, are the same as the SM. On the other hand, the effective 1-loop coupling to $\gamma\gamma$ and $Z\gamma$ receive contributions from the new charged gauge bosons, $V^\pm$ and $U^{\pm\pm}$, and the charged scalars.  With no other particles to decay to, the experimental signatures expected for the Higgs boson in 331 models should look very similar to the SM, but the channels related to photons decays.

As we pointed out in the previous section, the impact of the charged scalars on the branching ratio of the Higgs boson in two photons $BR_{331}(h\to\gamma\gamma)$ is negligible. We checked that for $m_V\gtrsim 100$ GeV, the contribution from scalars amount to less than 2\% for $100\gev$ charged vectors. As we will see, the preferred charged vector masses that fit the available data lies in the region $m_V > 150\gev$, so we can safely neglect the charged scalars in the calculations.

We are going to show in this section that a model with an extended gauge sector, as the 331 models, are able to explain the current observed Higgs branching ratios at the LHC within the current statistical errors.
We also emphasize that there exist minimal 331 constructions that possess a cold dark matter (CDM) candidate. This is a key feature for and experimentally well founded new physics model.

Although the minimal 331 model, as presented in Ref.~\cite{h2f331min}, does not contain a CDM candidate, it can be embedded in a larger gauge group, 341~\cite{341} at least, that has a neutral scalar which can be the lightest typical 341 particle~\footnote{This particle can be made stable by imposing a symmetry that transforms only the 341 fields which are singlet under the 331 symmetry. It is in this sense that we call it a typical 341 particle.}, and its mass may be varied such as to lead to a suitable CDM candidate~\cite{341R}.

Even more interesting, when the minimal 331 is supersymmetrized~\cite{susy331} (SUSY331), the lightest supersymmetric particle, generally a neutralino, is a good CDM candidate protected by R-parity, as in the Minimal Supersymmetric Standard Model (MSSM)~\cite{martin}. Considering the observation of an 125 GeV Higgs boson, however, the supersymmetric 331 presents an important advantage over the MSSM -- the upper bound on the Higgs boson mass lies comfortably above the measured mass when radiative corrections are considered~\cite{susy331,Rsusy331}, demanding less fine tuning (or none) in the parameters of the model, mainly the scalar top mass.

In both cases just discussed, it is reasonable to keep only the lightest particle, the CDM candidate, at low energy scale (some 10 to 100 GeV), while the remaining extra fields in the spectrum may be at the TeV scale or so, thus decoupling from the electroweak breaking regime and playing no role in the Higgs decay branching ratios.  Moreover, we will show  that if the branching ratios of the Higgs into the other SM particles are smaller than the predicted by the SM, then the invisible decay mode can be a natural way to decrease the branching fractions~\cite{farinaldo-etal}. From now, whenever we speak of the minimal 331 model we mean a 331 model with no DM candidate, while 331DM will refer to those models with a DM candidate (341 or SUSY331).

We define the ratio $\mu_{\gamma\gamma}$ between the branching fraction of a Higgs boson decaying into two photons of the 331 models and the SM as follows
\begin{equation}
\mu_{\gamma\gamma}=\frac{BR_{331}(h\to\gamma\gamma)}{BR_{sm}(h\to\gamma\gamma)}
\label{eq:rgg}
\end{equation}
and between the $s=Z,W,b,\tau$ branching ratios from the 331 and the Standard Model
\begin{equation}
\mu_{ss}=\frac{BR_{331}(h\to ss)}{BR_{sm}(h\to ss)}
\label{eq:rss}
\end{equation}
The SM Higgs boson widths and branching ratios were computed with the \texttt{HDECAY}~\cite{hdecay} program.
\begin{figure}
\centering
\includegraphics[scale=0.45]{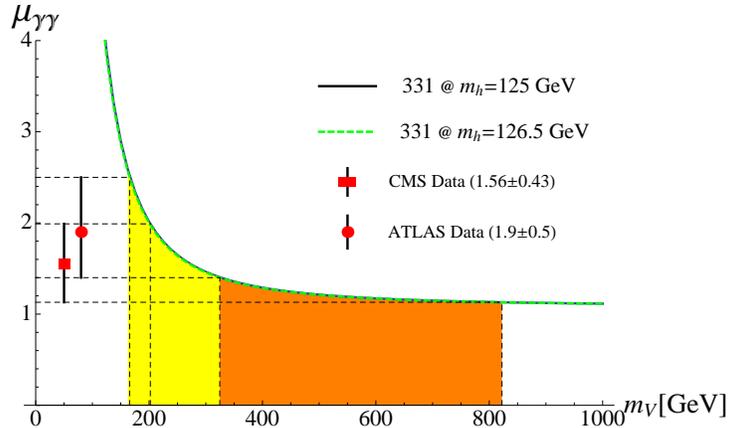}
\caption{The ratio $\mu_{\gamma\gamma}$ as a function of the common charged vector bosons masses $m_V$ for $m_h=126.5\gev$ (solid line) and $m_h=125\gev$ (dashed green line). We also show the $\mu_{\gamma\gamma}$ constraint from the latest CMS and ATLAS data points on the charged vector masses. The two lines are hardly distinguishable in this scale.}
\label{fig:rgg}
\end{figure}

We show $\mu_{\gamma\gamma}$ as a function of the $m_V$ masses in the Figure~(\ref{fig:rgg}). The solid line represents $\mu_{\gamma\gamma}$ from a minimal 331 model with $m_h=125\gev$, the central value from CMS, and the dashed line a $m_h=126.5\gev$ Higgs, the preferred value from ATLAS (almost indistinguishable from the solid line). In order to illustrate the experimental constraint from the recent LHC data on the Higgs search, we plot the $\mu^{CMS}_{\gamma\gamma}=(1.56\pm 0.46)$ and $\mu^{ATLAS}_{\gamma\gamma}=(1.9\pm 0.5)$ data points as measured by the CMS and ATLAS~\cite{CMS-ATLAS-sem2012} collaborations, respectively.

For the $1\sigma$ band variation $1.13<\mu^{CMS}_{\gamma\gamma}<1.99$, the $m_V$ masses lie in the range $(200\gev,\;825\gev)$, and in the range $(170\gev,\;320\gev)$ for $1.4<\mu^{ATLAS}_{\gamma\gamma}<2.4$ as can be seen in the Figure~(\ref{fig:rgg}). Of course, as the uncertainty in the data decreases these ranges will become narrower and a more precise prediction will be possible. Notwithstanding, these sub-TeV mass ranges seem to be well within the search reach of the 8 TeV LHC. Similar constraints follow from earlier ATLAS and CMS data.

\begin{figure}
\centering
\includegraphics[scale=0.4]{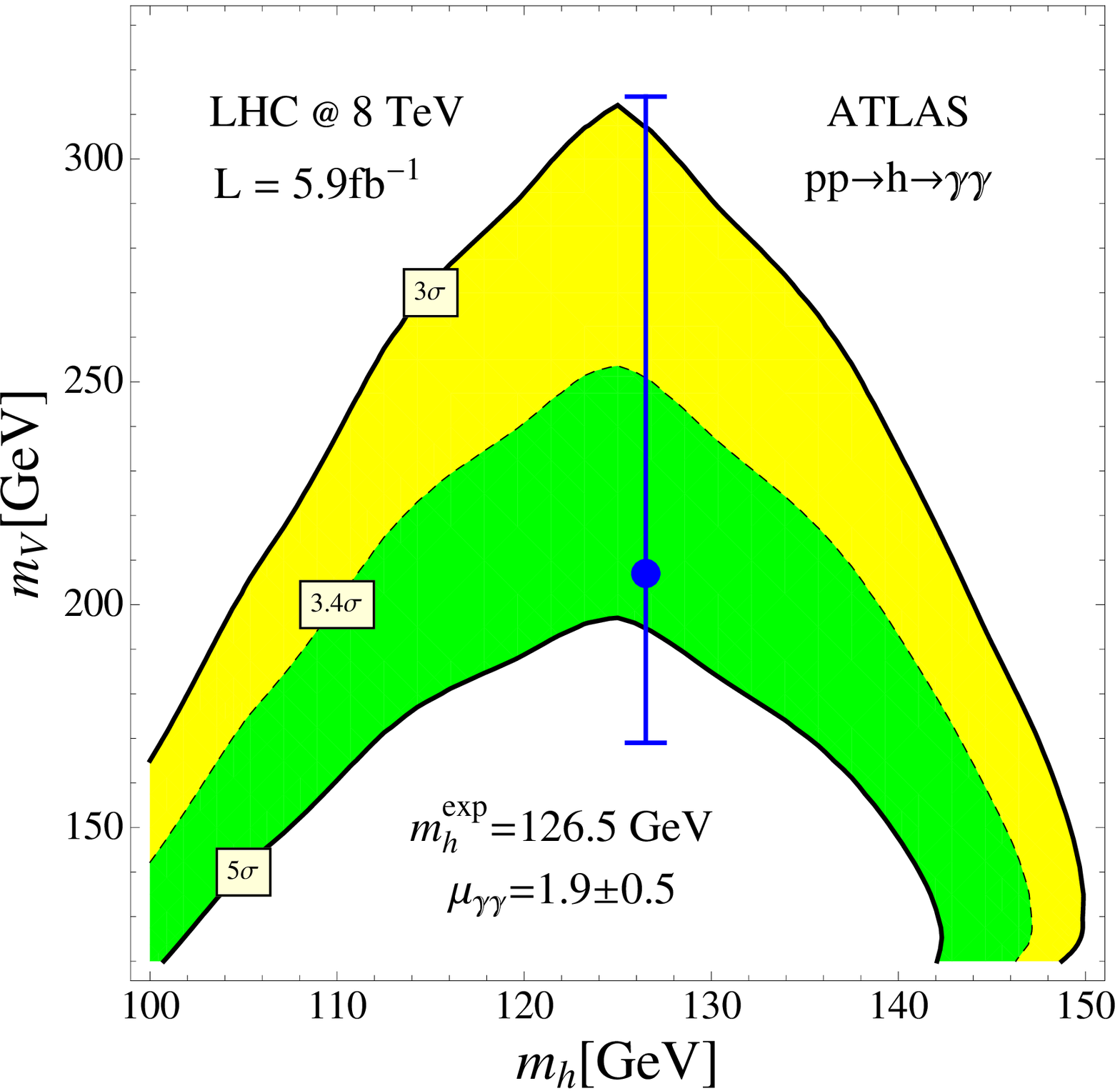}
\includegraphics[scale=0.4]{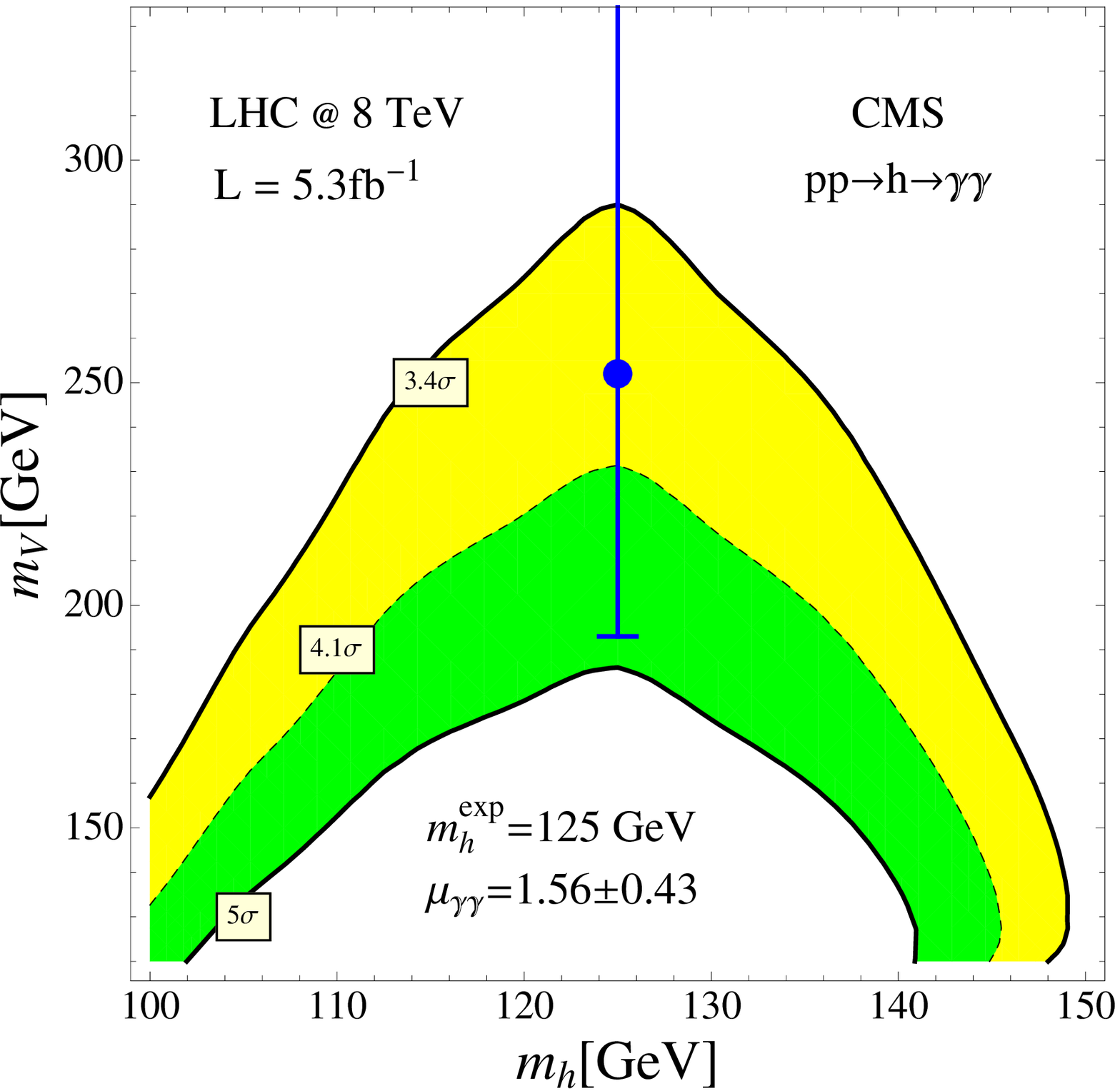}
\caption{In the right(left) panel we show the expected sensitivity of the LHC 8 TeV after 5.3(5.9) fb$^{-1}$ of collected data at the CMS(ATLAS) in the $gg\to h\to\gamma\gamma$ channel updated from Ref.~\cite{h2f331min}. The blue bars represent the mass constraint on the charged vectors from the minimal 331 model from the latest CMS(ATLAS) data on the signal strength $\mu_{\gamma\gamma}$.}
\label{fig:signif}
\end{figure}

In the analysis made in Ref.~\cite{h2f331min}, a Higgs boson with mass 125 GeV was found to give a $\sim 3\sigma$ signal at the LHC, after 5 fb$^{-1}$ of integrated luminosity have been accumulated, for new charged gauge bosons masses of 280 GeV. In the Figure~(\ref{fig:signif}) we update the expected significances at the 8 TeV LHC with 5.3 and 5.9 fb$^{-1}$ for CMS and ATLAS, respectively, for minimal 331 models. This picture will not change too much for 331DM models however~\footnote{In 341 models, the expected signal rate for the process $pp\to h\to\gamma\gamma$ is the same as the minimal 331. The SUSY331 gives additional contributions both to the Higgs boson production in gluon fusion, as the Higgs decay to photons. However, for a heavy SUSY spectrum the extra states would have a small impact on $\sigma \times BR(h\to\gamma\gamma)$.}. The blue points in the figure represents the solution for the vector boson mass obtained from $m_h=126.5(125)$ GeV and $\mu_{\gamma\gamma}=1.9(1.53)$, the experimental central values from CMS(ATLAS), while the blue bars represent the uncertainty in these data.

Note that expected significances are remarkably close to the observed significances of $3.4(4.1)\sigma$ from CMS(ATLAS)~\cite{CMS-ATLAS-sem2012} for the $h\to\gamma\gamma$ channel only. We point out, however, that the analysis made in Ref.~\cite{h2f331min} uses a less powerful statistics for the hypothesis test and somewhat different kinematic cuts compared to the experimental analysis. On the other hand, we calculated the main reducible and all the irreducible backgrounds at NLO accuracy, including single and double bremsstrahlung effects, and the Higgs production in gluon fusion at NLO QCD+EW. See Ref.~\cite{h2f331min} for more details.

There is a number of ways to confront the minimal 331 and 331DM model explanation to other candidate models. First of all, the new heavy 331 quarks do not couple to the Higgs boson as discussed in previous section. If the susy spectrum is heavy enough, the contributions from SUSY331 will also be negligible. So, the $hgg$ coupling is of SM size -- many other extensions give extra contributions to this coupling increasing the Higgs production cross section in gluon fusion. The $h\to Z\gamma$ is expected to change as well due the same new gauge bosons running in the loop.  Direct search for the new gauge bosons would be the ultimate test, once their masses would be of sub-TeV order and possibly accessible to the LHC.

As we pointed out before, the tree level couplings of the Higgs boson to the $Z,W,g$ bosons and to all the fermions are identical to the SM ones in 331 models. It means that if the Higgs boson decays to new states then all branching fractions to SM particles, dominated by tree level couplings, will be affected by the same factor. A few models realize this situation in a more natural fashion, for example, an spontaneously broken ${\cal N}=1$ SUSY with a  sgoldstino~\cite{bellazzini} is able to enhance the diphoton signal and keep the other branching ratios untouched at the cost of requiring either the wino or the bino to have a mass of the order of the gluino to avoid an overall enhancement due a larger $hgg$ coupling. Another example is a Higgs impostor, as the Randall-Sundrum radion proposed in~\cite{cheung}, where the $h\gamma\gamma$ and $hgg$ are enhanced due to trace anomaly. In these two examples, a reduction in the branching ratios to SM particles, but the photon, can be achieved increasing the $BR(h\to gg)$. Unfortunately, it is very difficult to detect this decay channel even in the Higgsstrahlung process due the overwhelming QCD backgrounds at hadron colliders.

If the Higgs decays to a pair of dark matter particles then, denoting such branching by $\alpha=BR(h\to invisible)$, the branching to a SM particle will be changed as follows
\begin{eqnarray}
BR_{331}(h\to ss) &=& \frac{\Gamma^{SM}_{ss}}{\tilde{\Gamma}_{SM}+\Gamma_{\gamma\gamma}^{331}(m_V)+\frac{\alpha}{1-\alpha}\left[\tilde{\Gamma}_{SM}+\Gamma_{\gamma\gamma}^{331}(m_V)\right]}\nonumber\\
&=&\frac{(1-\alpha)\Gamma^{SM}_{ss}}{\tilde{\Gamma}_{SM}+\Gamma_{\gamma\gamma}^{331}(m_V)}
\end{eqnarray}
In this formula, $s$ denotes all the SM particles but the photon and $\tilde{\Gamma}_{SM}=\Gamma_{tot}^{SM}-\Gamma_{\gamma\gamma}^{SM}$. The branching ratio to photons in 331 models is given similarly by
\begin{equation}
BR_{331}(h\to\gamma\gamma) = \frac{(1-\alpha)\Gamma_{\gamma\gamma}^{331}(m_V)}{\tilde{\Gamma}_{SM}+\Gamma_{\gamma\gamma}^{331}(m_V)}
\end{equation}
where $\Gamma_{\gamma\gamma}^{331}(m_V)$ is the partial width for a pair of photons as a function of the new gauge boson masses $m_V$.
\begin{figure}
\centering
\includegraphics[scale=0.5]{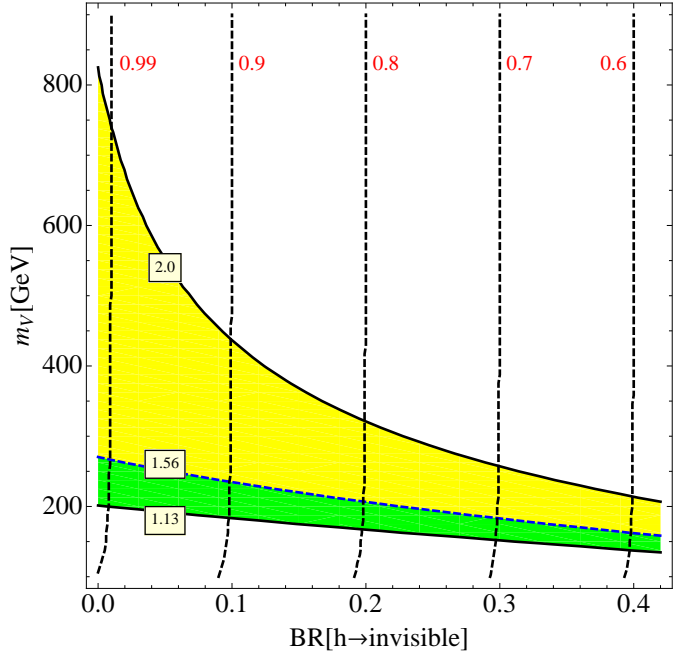}
\caption{The effect of an invisible decay channel in the ratio $\mu_{\gamma\gamma}$ in 331DM models. The green(yellow) band represents the downward(upward) $1\sigma$ mass constraint from the CMS data~\cite{CMS-ATLAS-sem2012}: $1.13<\mu^{CMS}_{\gamma\gamma}<2.0$. The dashed blue line represents the experimental central value. The almost vertical dashed lines show $\mu_{ss}$ as a function of the invisible branching ratio.}
\label{invisible}
\end{figure}

The effect of an invisible decay channel in the Higgs branching ratios into SM particles is to linearly decrease these branching ratios. If we want to keep $\mu_{\gamma\gamma}$ in the ballpark of the experimental CMS value, for example, we need lighter new gauge bosons. This can be seen in the Figure~(\ref{invisible}) where we show $\mu_{\gamma\gamma}$ and $\mu_{ss}$ in the $m_V$ {\it versus} $BR(h\to invisible)$. The blue dashed line shows the central experimental CMS value for $\mu_{\gamma\gamma}$ and the green(yellow) band the $1\sigma$ downward(upward) variation. The dashed vertical lines show
$\mu_{ss}$ in the $m_V$ {\it versus} $BR(h\to invisible)$ plane. Their values are quite insensitive to $m_V$ values, but decreases linearly with  $\alpha$.

\section{Confronting Higgs decays from 331 Models to the LHC and Tevatron data}

Despite the current data is compatible with the SM predictions within the current experimental errors, it has been shown that some sensitivity to new phenomena might be reached already combining all the released data from LHC and Tevatron collaborations~\cite{corbett-etal,giardino-etal,espinosa-etal,ellis-you,carmi-etal}. In fact, these works suggest that the couplings of the Higgs boson to the weak gauge bosons and fermions are compatible to the SM values, except for the photons possibly. Besides, a  branching ratio into invisible states can be accommodated~\cite{espinosa-etal,carmi-etal}. This situation can be naturally explained within the 331 models presented here.

A closer look at the data shows that many measurements indicate a smaller signal strength compared to the SM expectations~\footnote{See Ref.~\cite{corbett-etal} for a good compilation of the relevant experimental data}. In order to study the possibility of an invisible decay channel plus an enhanced branching ratio into pairs of photons within the 331 models, we performed a $\chi^2$ analysis using part of the available data from the LHC and Tevatron collaborations. For that aim we construct the following $\chi^2$ statistics with $m_V$ and $\alpha=BR(h\to invisible)$ as free parameters
\begin{equation}
\chi^2=\sum_{n=1}^{N_{exp}}\left[\sum_{s=Z,W,b,t}\left(\frac{\mu^{exp}_{n,ss}-\mu_{ss}(m_V,\alpha)}{\sigma^{exp}_{n,ss}}\right)^2+\left(\frac{\mu^{exp}_{n,\gamma\gamma}-\mu_{\gamma\gamma}(m_V,\alpha)}{\sigma^{exp}_{n,\gamma\gamma}}\right)^2\right]
\label{eq:chi2}
\end{equation}
where $N_{exp}$ ranges from $1$~\cite{atlasexp7}, $2$~\cite{cmsexp7}, $3$~\cite{CMS-ATLAS-sem2012}, $4$~\cite{tevexp}, to $5$~\cite{CMS-ATLAS-sem2012}. We quote these data in Table~(\ref{tab}).
%
% contours figure
The quoted experimental errors are asymmetric but we take the average of the upper and lower variances to compute $\sigma^{exp}_{ss}$, $\sigma^{exp}_{\gamma\gamma}$, and the $\chi^2$ statistics. The experimental collaborations do not provide us the correlation matrices, so we can take into account neither the possible correlations among the experimental data sets of the collaborations nor between the 7 and 8 TeV runs. We also do not take systematic uncertainties (as the theory errors on the production cross sections) into account, so our results should be taken as a rough estimate of the best $m_V$ masses and invisible branching ratio which fit the data therefore.
\begin{table}
\begin{center}
\begin{tabular}{c|cccccc}
\hline
\hline
$N_{exp}$  & $\mu^{exp}_{\gamma\gamma}$ & $\mu^{exp}_{ZZ^*}$ & $\mu^{exp}_{WW^*}$ & $\mu^{exp}_{bb}$ & $\mu^{exp}_{\tau\tau}$ \\
\hline
(1) ATLAS 7 TeV & $1.6\pm 0.81$ & $1.4\pm 0.80$ & $0.5\pm 0.7$ & $0.5\pm 2.05$ & $0.2\pm 1.80$\\
\hline
(2) CMS 7 TeV & $1.5\pm 1.05$ & $0.6\pm 0.77$ & $0.4\pm 0.6$ & $1.2\pm 1.96$ & $0.6\pm 1.15$ \\
\hline
(3) CMS 7+8 TeV & $1.56\pm 0.43$ & $0.7\pm 0.44$ & $0.6\pm 0.4$ & $0.12\pm 0.70$ & $-0.18\pm 0.75$ \\
\hline
(4) ATLAS 7+8 TeV & $1.9\pm 0.5$ & $1.3\pm 0.6$ & -- & -- & -- \\
\hline
(5) CDF and D0 & $3.6\pm 2.76$ & -- & $0.32\pm 0.83$ & $1.97\pm 0.71$ & --\\
\hline
\hline
\end{tabular}
\end{center}
\caption{Experimental data used in the fitting procedure. The symmetric errors are computed from the actual asymmetric experimental errors by averaging their variances.}
\label{tab}
\end{table}

\begin{figure}
\centering
\includegraphics[scale=0.5]{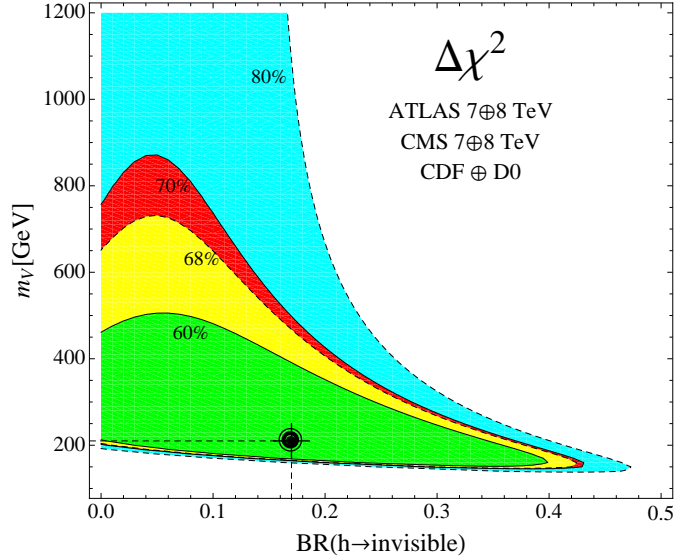}
\caption{The $\Delta\chi^2$ contours in the plane $m_V$ {\it versus} $\alpha=BR[h\to invisible]$ corresponding to the Confidence Levels of 60\%, 68\%, 70\%, and 80\% computed from the table~(\ref{tab}). The plotted point locates the best fitted parameters from the data.}
\label{fig:chi2}
\end{figure}

After computing the global $\chi^2$ we determine the minimum $\chi^2_{min}$ and plot the modified statistics $\Delta\chi^2=\chi^2-\chi^2_{min}$ in Figure~(\ref{fig:chi2}) where the 60\%, 68\%, 70\%, and 80\% C.L. contours are shown in the $m_V$ {\it versus} $BR(h\to \hbox{invisible})$ plane.  The contour values, for a given confidence level $\lambda$, are calculated from the inverse cumulative distribution function (CDF)of the $\chi^2_{n_{dof}}$ probability density for $n_{dof}=2$ degrees of freedom: $\Delta\chi^2_\lambda=\hbox{CDF}_{\chi^2}(1-\lambda,n_{dof})$. The 68\% C.L. correponds to $\Delta\chi^2_{0.68}=2.30$, for example~\cite{gcowan}.
From our fitting procedure we found $\chi^2_{min}=13.72$ for 20 data points. The $\chi^2$ for the SM hypothesis is $17.23$ which is inside the 84\% region in the $m_V$ {\it versus} $BR(h\to invisible)$ plane and agrees well with the number found in~\cite{corbett-etal}, for example.

The best fitted point in this parameters space is $\left[m_V,BR(h\to \hbox{invisible})\right]=[212.5\gev,0.17]$, and the best theoretical signal strengths corresponding to these values are $\mu_{\gamma\gamma}=1.57$ and $\mu_{ss}=0.83$ which agree reasonably well with the CMS and ATLAS values~\cite{CMS-ATLAS-sem2012}, while the 68\% C.L. intervals are $(164.4\gev,471.7\gev)$ and $1.02<\mu_{\gamma\gamma}<2.1$.

We point out that a 17\% branching ratio in invisible decays is in fairly good agreement with  similar analysis made in Refs.~\cite{espinosa-etal,carmi-etal}. Moreover, a general fit of the Higgs couplings~\cite{corbett-etal} found that the present data favor a 55\% smaller higgs production rate in gluon fusion compared to the SM rate. This is a consequence of the deficits found in 13 out of the 29 data points used in the fitting procedure in that work. If we suppose this is not a effect of a fainter $hgg$ coupling, an invisible decay is the best alternative to a global decrease in the observed Higgs boson branching ratios.

The fitting is dominated by the $\gamma\gamma$, $ZZ^*$, and $WW^*$ data from ATLAS and CMS which quote the smaller experimental errors. Despite a somewhat large branching to invisible decays is preferred by the data in order to fit the dominating $ZZ,WW$ channels, a vanishing $BR(h\to \hbox{invisible})$ is within the 68\% confidence interval. The SM point lies in $BR(h\to \hbox{invisible})=0$ line for large $m_V$. On the other hand, a SM branching into photons seems less favored by the current data.

Supposing that the Higgs boson decays to SM particles exclusively we fit $BR(h\to\gamma\gamma)$ to the data as a function of $m_V$ obtaining $m_V=267.5\gev$ as the best fitted masses and $\mu_{\gamma\gamma}=1.59$ as the best signal strength value fitted from the data, for a Higgs boson mass of $126\gev$.

\section{Conclusions and Outlook}

The goal of discovering the Higgs boson has finally been achieved at the LHC and corroborated by an evidence signal at the Tevatron. In fact, telling the resonance is a Higgs boson is nothing but a pretty good guess in this moment. The next logical effort, both experimental and theoretical, is to study the new particle's properties in order to confirm or not its role in the electroweak symmetry breaking.

Some beyond SM models are able to explain the current state of affairs concerning the branching ratios of the hypothesized Higgs, however, it is not generically easy to adjust the couplings of the Higgs to SM particles in order to fit the current data. One of the most economical alternatives to this scenario would be to keep all SM couplings untouched and add new states to which the Higgs could decay to, decreasing the branching ratios globally at the same time these new states enhance the 1-loop coupling to photons.

A few models realize this situation in a more natural fashion including the classes of 331 models considered in this work, in special,  the versions presenting a Cold Dark Matter particle. In these realizations, the Higgs boson may decay to CDM particles and become invisible to the detectors while the couplings to the rest of the spectrum would look SM.

In this work we show that the minimal 331 model, and versions presenting a dark matter candidate, can fit the current data within the experimental errors. We performed a $\chi^2$ analysis using the publicly available data (with unknown correlations) and found that the proposed models with charged vector masses of $212.5\gev$ and a  branching ratio to invisible states of  17\% are the preferred parameters from the fitting procedure. Furthermore, given the large errors, larger masses and smaller branching ratios to invisible states cannot be excluded. The preferred signal strength to photons from the fitting procedure is $\mu_{\gamma\gamma}=1.57$, which agrees well with the LHC data. If no DM is present in Higgs decays, our analysis finds $\mu_{\gamma\gamma}=1.59$ and $m_V=262.5\gev$ as the best fitted mass.

A general prediction of 331 models with a Cold Dark Matter candidate is to globally decrease the branching ratios to SM particles. Given the observation of deficits in many experimental data, and allowing an invisible decay channel, as, in fact, is predicted by supersymmetric 331 models and minimal 331 models embedded in larger groups, as the 341, our $\chi^ 2$ analysis shows that the best global signal strength to all SM particles, but the photon, fitted to the data is $\mu_{ss}=0.83$. Comparing this to the CMS global signal strength $\sigma / \sigma_{SM}=0.80\pm 0.22$, we may say that the 331 explanation to the Higgs branching ratios and the signal strength in diphoton channel is robust within the current experimental errors.

Whatever the direction the experimental picture evolves, these models are able to describe a scenario with either decreased branching ratios into $Z,W$ and $b,\tau$ pairs or SM branching fractions, and either a photophilic or a SM Higgs concerning the branching to photons.  Nevertheless, the charged vector bosons masses preferred by the data suggest that a direct search for new gauge bosons from the models considered here is well within the reach of the LHC. We also point out that if it could be possible to compare the yields of the associated process $pp\to hZ(W)\to ggZ(W)$ with the SM prediction this would help to distinguish between the 331 models and other models that globally change the branching ratios of the Higgs into SM particles. Further signals for distinction of the extended electroweak models we deal with here would be, for example, the ones involving a peculiar kind of fermionic leptoquark $J_3$~\cite{leptoquarks331} whose decay proceeds through $J_3\rightarrow b\,U^{++}\rightarrow\ b+l^+l^+$.

\acknowledgments
This work was supported by Conselho Nacional de Pesquisa e
Desenvolvimento Cient\'{i}fico - CNPq, Coordena\c{c}\~ao de Aperfei\c{c}oamento Pessoal de N\'ivel Superior - CAPES, and Funda\c{c}\~ao de Amparo \`a Pesquisa do Estado de S\~ao Paulo - FAPESP.

%%%%%%%%%%%%%%%%%%%%%%%%%%%%%%%%%%%%%%%%%%%%%%

%%%%%%%%%%%%%%%%%%%%%%%%%%%%%%%%%%%%%%%%%%%%%%%%%%%%%%%%%%%%%%%%%%%%%%%%%%%%%%%%%%%%%%%%%%%

\end{document}